# Adaptive machine learning strategies for network calibration of IoT smart air quality monitoring devices

Saverio De Vito[a]∗, Girolamo Di Francia[a], Elena Esposito[a], Sergio Ferlito[a], Fabrizio Formisano[a] and Ettore Massera[a]

[a]ENEA – DTE-FSN-SAFS, P.le E. Fermi, 1, 80055 Portici (NA), Italy

## Abstract

Air Quality Multi-sensors Systems (AQMS) are IoT devices based on low cost chemical microsensors array that recently have showed capable to provide relatively accurate air pollutant quantitative estimations. Their availability permits to deploy pervasive Air Quality Monitoring (AQM) networks that will solve the geographical sparseness issue that affect the current network of AQ Regulatory Monitoring Systems (AQRMS). Unfortunately their accuracy have shown limited in long term field deployments due to negative influence of several technological issues including sensors poisoning or ageing, non target gas interference, lack of fabrication repeatability, etc. Seasonal changes in probability distribution of priors, observables and hidden context variables (i.e. non observable interferents) challenge field data driven calibration models which short to mid term performances recently rose to the attention of Urban authorithies and monitoring agencies. In this work, we address this non stationary framework with adaptive learning strategies in order to prolong the validity of multisensors calibration models enabling continuous learning. Relevant parameters influence in different network and note-to-node recalibration scenario is analyzed. Results are hence useful for pervasive deployment aimed to permanent high resolution AQ mapping in urban scenarios as well as for the use of AQMS as AQRMS backup systems providing data when AQRMS data are unavailable due to faults or scheduled mainteinance.

∗ Corresponding author: Saverio De Vito, Ph.D.. Tel.: +39-081-772-3264; e-mail: saverio.devito@enea.it

# 1. Introduction

Low cost chemical multisensors devices are on the verge of revolutionize Air Quality (AQ) monitoring networks bringing in the possibility of personal exposure assessment while largely improving the pervasivity of current regulatory grade analyzers network. While regulatory monitoring network will continue to represent the backbone for providing the highest quality AQ data, several cities are now devising new AQ monitoring network for reaching unprecedented resolution on air quality levels assessments. In particular, European Union determination in improving, starting at cities levels, the AQ throughout its territory have fueled 5 innovation projects based on high density deployments of intelligent AQM systems (UIA AQ website, 2020). However, this can only be achieved by trading off density of measurements with a loss of accuracy performances at single device level (Lewis et al., 2016). Despite the significant research efforts in the last five years in which field studies have shown that their performance have gradually reached surprising levels, they still have to meet the strict uncertainty requirements that would qualify them for use as indicative measurement systems (Borrego et al., 2016; EU AQ Directive, 2008). Despite their higher relative costs with respect to Metal Oxide sensor devices, Electrochemical sensors are being validated as the most promising in long term operative deployments of networked AQMS. Low cost sensors are actually affected by a number of technological issues including non linearity and interferences from known (or even unknown) non-target gases and environmental conditions (Cordero et al., 2018; Cross et al., 2018;). Furthermore, chemical sensors response to their target gas and non-target gas as well as to environmental interferents may be non-linear. A useful example is temperature influence in electrochemical low cost sensor devices response (Alphasense $NO_2$ sensor datasheet, 2020). Fabrication process induced variance determines sensitivity and zero gas response to significantly vary from one device to another resulting in the need for separately calibrating each and all chemical (multi-)sensors device. Finally, the response function may vary in time for known and unknown effects like sensor poisoning, due to long term exposure to significant concentrations of the target gas, or ageing (Marco and Gutierrez-Galvez, 2012).

Calibration of these systems is hence crucial for assessing and validating their performance so that they could be confidently used for obtaining meaningful information on AQ in cities (Borrego et al., 2018). Unfortunately, the number of conditions to be reproduced in a controlled atmosphere setup to fully understand a chemical multisensory response is so high to currently impede to resort to (lengthy) lab based calibration procedures. Field data driven multivariate calibration, using AQRMS as ground truth, may significantly reduce the impacts of known interferents by learning models that can remove their non linear impacts on sensor response. Machine learning model have shown to be best candidates for on board or remote calibration of chemical sensors array operating in air quality monitoring scenarios (Borrego et al, 2018; Esposito et al., 2016). However unknown interferents can play a hidden context role, negatively affecting model robustness to changing conditions. Atmospheric field conditions are in facts non-stationary and machine learning models validity may fade in time due to seasonal changes in anthropogenic emissions and environmental conditions and, of course, sensors drift (Masey et al., 2018, De Vito et al, 2012). This setting is known in machine learning community as *learning in dynamic environments*, and can be addressed with different methodologies including frequent recalibration, fusion of seasonal classifiers, semi-supervised learning, etc. appropriately reacting to change detection (Ditzler et al., 2015). The above mentioned peculiarity of the chemical sensing field, and particularly fabrication variance, sensor ageing and limited lifespan of chemical sensors units, needs ad hoc approaches. In facts, recalibration procedures may significantly reduce the loss of accuracy due to overall concept drift but high quality labeled data are usually available only at initial calibration time. The high cost of re-collecting instruments and awaiting for the re-calibration time is however a braking factor for commercial enterprises and customers. Hence, we need to resort to ad-hoc procedures that exploit any event, whether opportunistic or pre-scheduled, to co-locate, at least briefly, one low accuracy device with one that can improve its performance based on accurate target concentrations readings. This approach is usually referred as node to node or rendez-vous based (re)calibration. Hasenfratz et al. pioneered the use of online algorithms for rendez-vous based recalibration in networks of air quality monitors (Hasenfratz et al., 2012). They used a fixed periodical recalibration scheme with nodes receiving regular 40mins periodic updates to their polynomial univariate calibration parameters. Three update models were compared among which one weighting past calibration tuples actually implementing a tunable forgetting mechanism aiming to deal with sensor ageing. Only sensors with recent calibration are used as information sources for node to node (N2N) calibration.  Field results using a two collocated $O_3$ MOX sensor and simulated replicas indicates the significant improvement of Mean Absolute Error (MAE) performances due to on-the-fly N2N calibration over a 6 weeks long deployment. Arfire et al., have exploited OpenSense dataset to statistically determine a fixed but realistic opportunistic rendezvous calibration scheme for public service bus mounted AQMS based on bus routes (Arfire et al., 2015). Using a ten months colocation experiments they used a regulatory grade instrument to calibrate a CO sensor device correcting for temperature interferents using several calibration function models. They did not compare the performances with a classic offline initial calibration scheme, which they did not exploit, but shown the positive impact of rendezvous calibration whenever the incoming data stream was tuned to the model complexity.  Only new samples are used to generate updated calibration actually retaining no memory of previous calibration parameters. Kizel et al., theorethically and experimentally studied the propagation of errors during multi-hop node to node univariate calibration procedures in AQMS networks (Kizel et al., 2018). They have shown the possibility to guarantee the error level to be kept in a preset range by accurately design the length of the calibration chain. Focusing on $O_3$ and NO they experimented a fixed recalibration scheme with a period of 3 weeks and a colocation length of one week during different scenarios over a maximum of  3 months. Optimal results where obtained keeping the multi-hop chain at a length <= 2 hops.  Another approach could not rely on physical co-location but rather on remote information exchange. In fact, it could be based on the continuous or periodical availability of a stream of high accuracy target gas concentration estimations projected at multisensors location by a geostatistical AQ mapping algorithm or model exploiting the knowledge arising by the neighboring network nodes readings. This approach can be indicated with the term network calibration (NC). Miskell et al. recently shown the results obtained by a simple but effective procedure (Miskell et al., 2018). One pioneering example of this strategy was already described by Tsujita et al., which updated their networked $NO_2$ sensors baseline response parameter using samples featuring very low concentrations expected at night during particular weather conditions (Tsujita et al, 2005). Network calibration strategies usually target fixed stations monitoring AQ parameters with time resolution in the hourly range usually obtained by averaging minute resolution data. A special case study is represented by the use of a low cost multisensors device as a backup station that is continuously co-located with a regulatory grade analyzers station for the purpose of providing backup data when the conventional analyzers are off-line. AQ Regulatory Monitoring Stations (AQRMS), in facts, suffer from periodical failures and are subjected to frequent maintenance operations that put them off-line. This causes data to be missing during several time periods along the year hampering the validation of short and long term assessments on AQ on urban and regional scale. Low cost multisensors device may hence allow to gracefully degrade the performance but still obtaining data in this unwanted but unfortunately frequent condition (De Vito et al., 2019).

None of these approaches have been tested for long term deployments (e.g. >1 year) while they mostly focus on N2N and univariate calibrations schemes. As mentioned before, multivariate calibration is crucial to exploit the information gathering potential of a chemical sensor array and reduce interferents driven error. A comprehensive and experimental exploration of the impact of frequency and amount of information exchange in network calibration approaches is still lacking. In this paper, we will therefore focus our efforts to test adaptive multivariate machine learning strategies in the non-stationary context of field operating air quality multisensors devices over long term deployments. Using data coming from a running co-location experiment now encompassing 18 months, we will test adaptive and incremental learning strategies to improve the accuracy of an initially calibrated low cost sensor device targeted to CO, $NO_2$ and $O_3$ pollutant estimations. Different periodic calibration schemes are tested to clarify the performance dependence on recalibration period and number of updated calibration tuples. The results will help to improve our knowledge on the performance obtainable by backup monitor nodes as well as network or node-to-node continuous recalibration strategies.

## 2. Experimental settings

### 2.1. The IoT Air Quality multi-sensors node architecture

In this study, we rely on the use of the ENEA MONICA AQMS, designed for cooperative mobile air quality quantitative sensing operations. MONICA device is based on electrochemical sensors array using Alphasense™ A4 class sensor units, respectively targeted to Carbon Monoxide, Nitrogen Dioxide and Ozone. Relative Humidity (RH) and Temperature sensors complete the sensing array. The sensors analog front-end is provided by the same company and allows to connect sensors to an ARM microcontroller based ST Microelectronic Nucleo board. The latter captures and digitalize the two relevant sensors terminal voltages of each sensors, namely the working (WE) and auxiliary electrodes (AE), along with the temperature and RH readings from a SHT11 sensor device. Raw sensors data are captured by a 10 bit ADC and transmitted via a Bluetooth serial interface to a Raspberry Pi Mod. 3 + based datasink with Raspbian OS providing for local storage and WAN connectivity services through a 3G connection. Data is captured at 10 samples/minute rate. At remote side, an ad-hoc IoT backend architecture relying on a contained NodeJs REST APIs server and MongoDB provides data inception, device management, storage, preprocessing and map based visualization functionalities. In this study, downloaded data is averaged on an hourly based so to be compared with hourly data outputted by reference instruments.

### 2.2. Recording the dataset

A MONICA device has been located, since April 2018, in Via Argine (Naples, Italy) along a main road connecting Naples to several towns along the Mt. Vesuvius area in co-location with regulatory grade analyzers station.

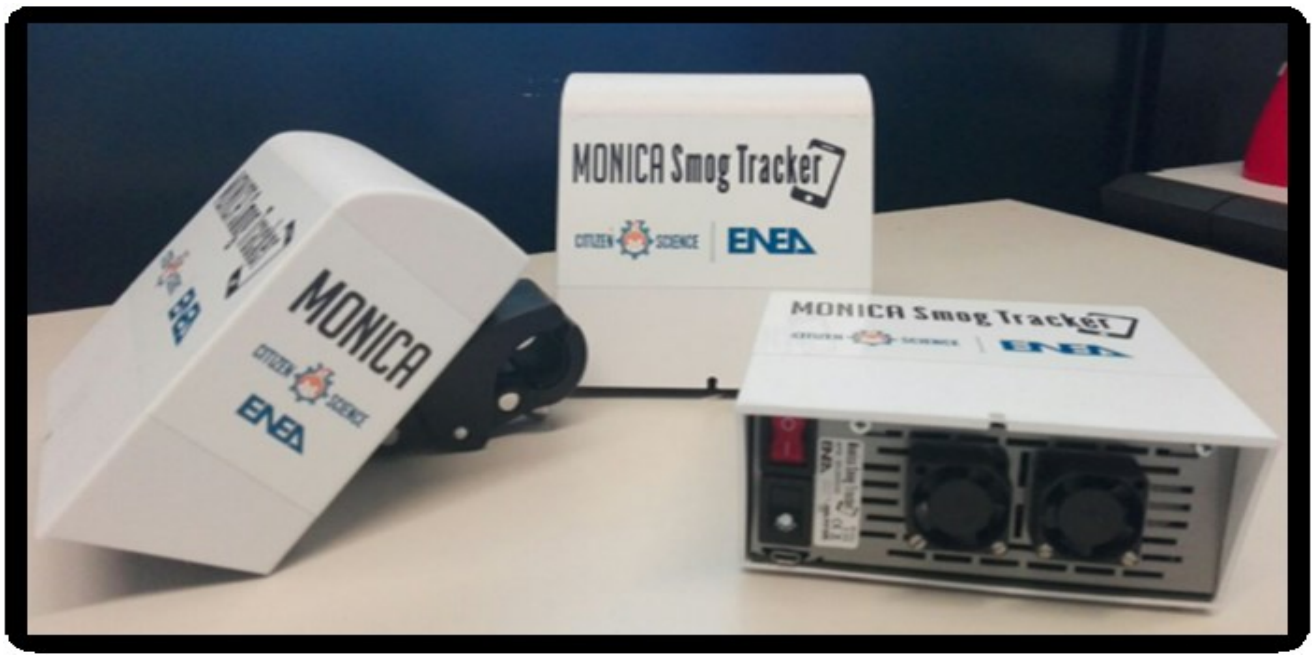

**Fig. 1:** Three MONICA devices in their outdoor casing (up).

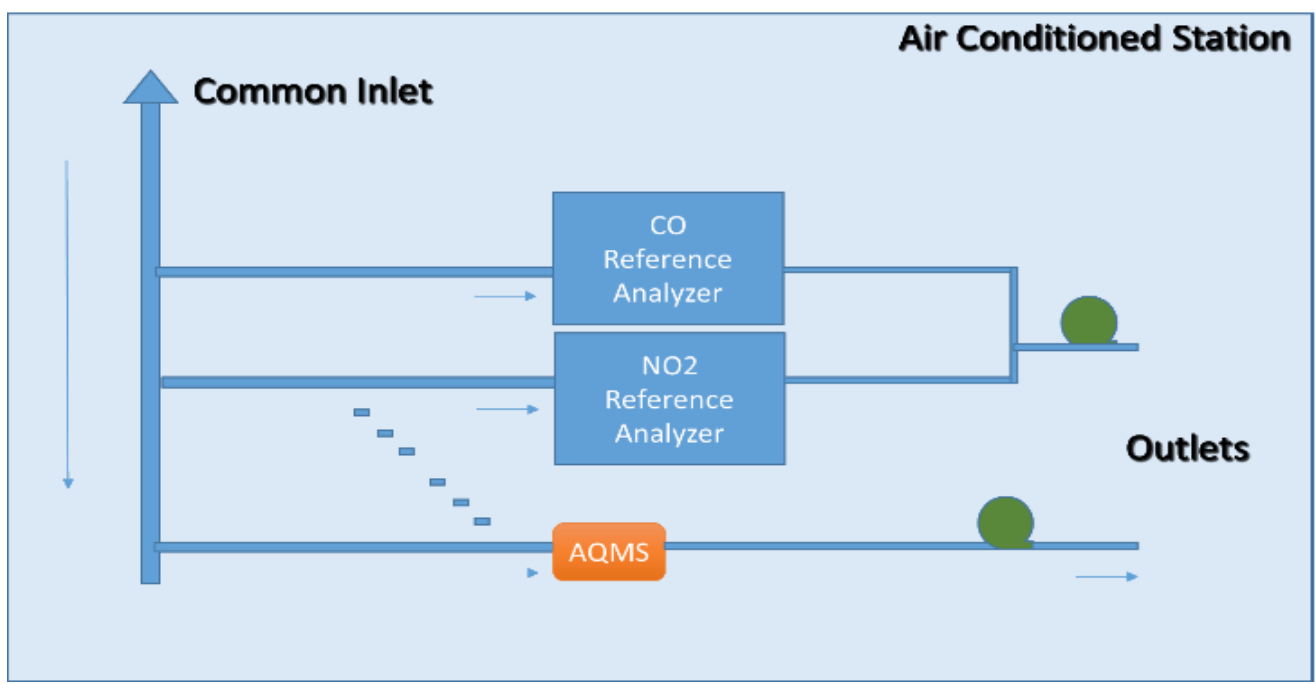

**Figure 2:** The connection schema showing the MONICA device as fed by the common inlet of the ARPAC station.

The latter is actually operated by the Regional Environmental Protection Agency of the Campania Region (ARPAC) and is part of the official regulatory AQ monitoring network, as such it is subjected to regular maintenance activities. Its deployment design includes a

Teledyne™ 300 CO analyzer, a Teledyne T200 NO2 chemiluminescence analyzer, a BTEX analyzer and Particulate matter (PM2.5 plus PM10) analyzer. Differently from other co-location studies, our AQMS inlet is directly connected on the main common inlet of the AQRMS (see Fig 1). The common inlet ensure that same air flow reaches all the regulatory instrumentation and the MONICA device. The setup is actually designed to limiting any interference coming from differences in the real time analyzed air usually found in the common co-location deployments. The recorded dataset consists of 6864 Hourly samples recorded from April 2018 to July 2019. Notwithstanding the presence of an Air conditioning (AC) unit, the inner of the reference station underwent significant temperature oscillations (see Fig. 3) which peaks occasionally reached 38°C in July 2018, due to incorrect set-point of the system. Specifically, the dataset contains the hourly averaged data from the device, i.e. Working Electrode (WE) and Auxiliary Electrode (AE) raw sensors readings (mV) for $NO_2$, CO, $O_3$ plus Temperature (°C), Humidity (%), joined to hourly averaged data from ARPAC reference analyzer for $NO_2$ (ppb), CO (ppm). In particular, the Auxiliary electrode readings may be used to partially correct for temperature interference affecting these sensors WE. It is, in fact, a twin electrode which surface never comes in contact with the pollutant while still being influenced by temperature. Due to their particular geometry and manufacturing difficulties, the effect of temperature on the different electrodes readings are different and temperature still affect their difference (WE-AE) which by the way remains the manufacturer adviced best input signal for the univariate calibration of these sensors (see Sensors Datasheet in reference section ). The results obtained in the calibration procedure section will focus on $NO_2$ while $O_3$ reference data were not available. In Fig. 4, monthly box plot of $NO_2$ and CO target gas true concentration behavior along the entire dataset are shown. In Fig. 5, $NO_2$ and CO time series are shown exhibiting the presence of holes in the data, due to failure in the BT connectivity for undocumented bugs in BT libraries for Raspbian. The dataset has been preprocessed, analyzing the missing values, detecting the possible outlier's while a correlation analysis has been carried out. Specifically, for or the outlier's detection, DBSCAN algorithm has been applied removing 72 anomalies in the entire dataset. This slightly improved the correlation between the target gas data and the sensors data. Fig. 6 show the monthly probability distribution of target and primary NO2 targeted sensor reading as corrected for temperature (AE-WE) alongside with their correlation factor.

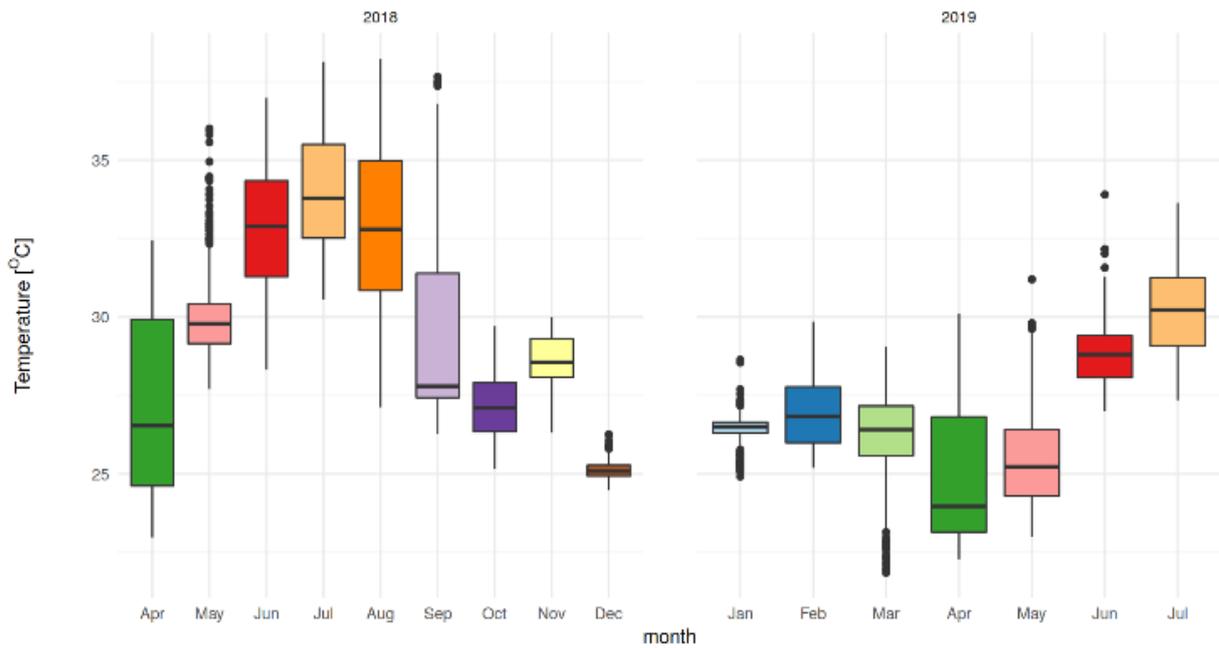

**Figure 3:** Monthly temperatures boxplot along the dataset. Outliers are depicted as single dots.

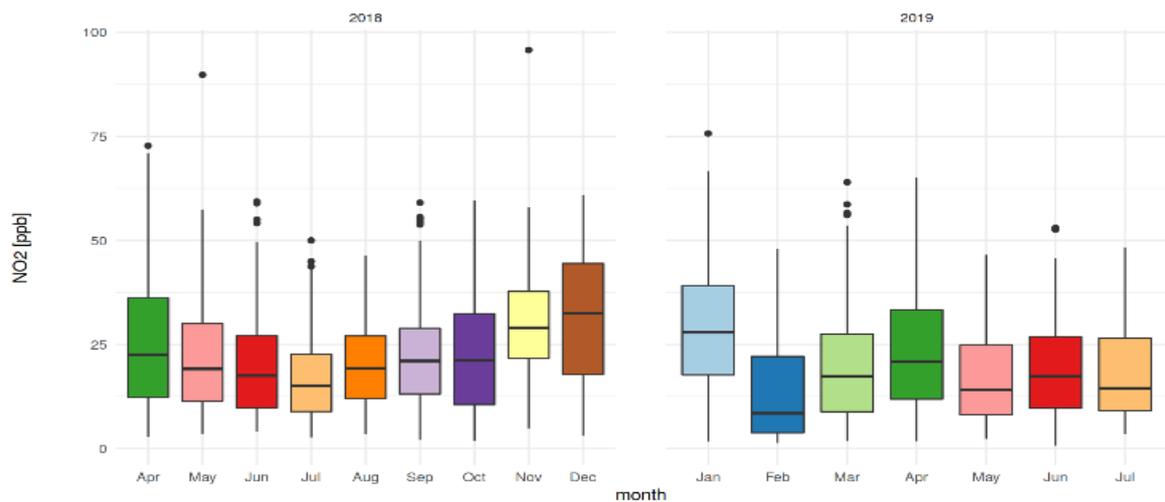

**Fig. 4:** Monthly box plot of $NO_2$ (ppb) reference concentration behavior along the data set. Outliers are depicted as single dots.

## 2.3. Calibration strategies

As above mentioned, chemical sensors arrays raw data needs to be processed by a calibration function to accurately estimate target gases concentration taking care of non linearities and interferents. Generally, a static calibration function can be expressed by:

$$C_{pollutant} = f(x_t, x_1, \ldots x_k, w) + g \quad (1)$$

Where $x_i$ are the observed sensor readings, $w$ is the function $f$ parameter vector values and g the unknown error term. In fact, information about both target ($x_t$) and known non target interferents gases ($x_1, \ldots x_k$) helps to build a more complete sensor response model.

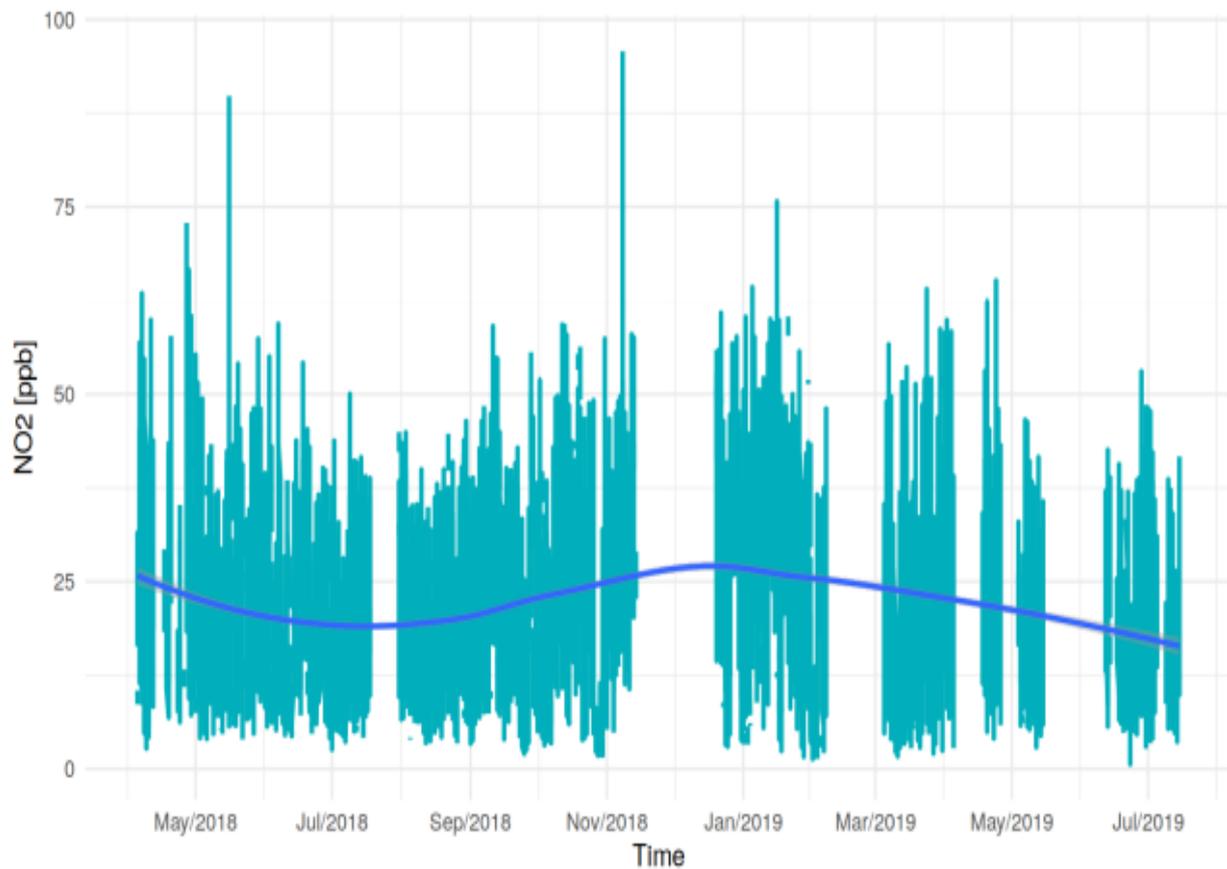

**Fig. 5:** NO$_2$ (ppb) ground truth time series as fitted with LOESS interpolation function.

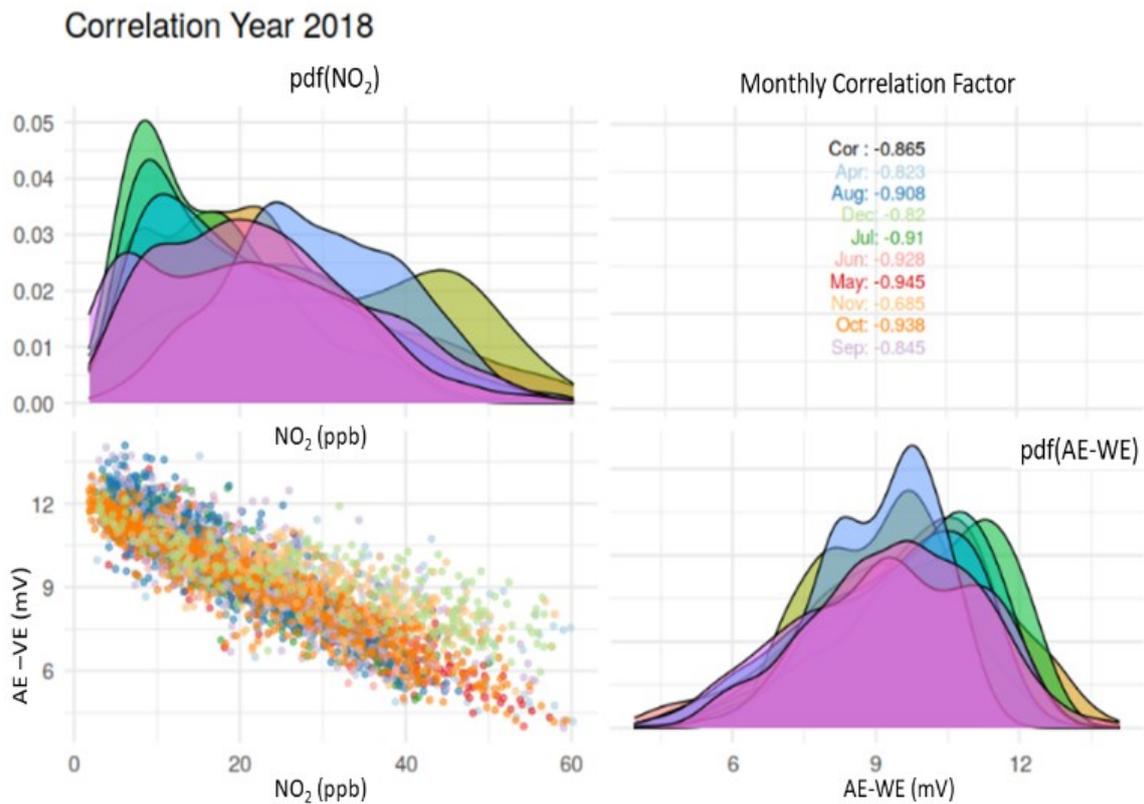

**Figure 6:** Pearson Correlation Coefficients between $NO_2$ target gas concentration and $NO_2$ raw sensor readings (AE-WE difference) as computed for each month in 2018 (upper right corner), monthly empirical *pdf* (upper left corner for $NO_2$ and lower right for AE-WE) and scatter plot among the above variables (lower left corner). Different colors are associated to different monhts. Significant, seasonal bias and sensitivity changes are evident. We advise the user to refer to the online version of this paper for best readability.

The number of observables can be completed by using past recordings of the inputs and outputs using tapped delay lines, aiming to cope with sensors dynamic and uncorrelated noise affecting these non linear time variant dynamic systems (Esposito et al., 2016). By using field recorded data, on field calibration methodology builds training set that actually reflect field operative conditions in terms of atmosphere composition and environmental settings. However, this condition may be altered by the emergence, during time, of concept drifts, forcing the specific calibration function to work outside the boundaries of its training manifold thus hampering the overall quality of the results.

Specifically, field calibration algorithms are challenged by changes in the joint probability distribution of sensors responses and their forcers concentrations:

$$P(x_i, C, \tilde{C}, E) \quad (2)$$

Actually, sensors ageing and pollution affects sensors sensitivity and baseline response, while seasonal changes in environmental and anthropic impacts affect target gas C concentration, known or unknown non-target interferents gases $\tilde{C}$ concentrations as well as temperature and humidity values (E) that act as an hidden context eventually modifying the conditional probability distribution:

$$P(x_i|C) \quad (3)$$

By explicitly taking care of relevant environmental variables values $e_i$, calibration functions (see eq. 1) can be trained to correct for their influence:

$$C_{pollutant} = f(x_t, x_1, ..x_k, e_i, w) + h \quad (4)$$

but, again, seasonal changes in their distribution may cause the learnt model to become incomplete. Non-observables interference **h** measurement is, by definition, completely beyond our grasp.

Summarizing, changes in the inherent sensor model or forcers distribution invalidate the learnt calibration function by respectively incoherence or incompleteness. Should new labeled samples become available, online learning strategies, either of incremental or adaptive nature, could be used to correct for these issues. However, catastrophic forgetting or interference effects may set in. To this purpose, we selected two algorithms namely standard shallow, feed forward, neural network (SNN) and an Extreme Learning Machine (ELM) to experiment, respectively, the two methodologies (online incremental re-training and updating).

In particular, shallow neural networks have already proven very effective for in field calibration of AQ multisensors systems while ELMs have barely been used in such scenarios. ELMs are simple feed forward architectures which exploit a random feature expansion in the first layer and a fully connected linear output layer. The underlying idea has been explored since decades and is one of the key basis of compressive sensing (Huang et al., 2006). They seems to share architectural similarities with mammals brain and bear fast computation advantages while providing good results especially when small dataset are concerned, when compared with best of class algorithms.

In this paper, we focused on NO2 concentration estimation problem using working electrode (WE) and auxiliary electrode (AE) sensors data of NO2, O3 and CO sensors plus temperature and humidity data as inputs for the two ML components:

$$C_{NO2} = f(WE_{NO2}, AE_{NO2}, WE_{CO}, AE_{CO}, WE_{O3}, AE_{O3}, Temp, RH)$$

Exploiting previous findings in similar scenarios (see De Vito et al., 2018), the number of hidden neurons for both architectures have been preset to lay in a fixed range. The SNN architecture was empirically equipped with [3,5,7] standard sigmoidal tangent neurons units in its sole hidden layer, while the final layer was equipped with a linear output layer. Automatic Bayesian Regularization (ABS) was used as training algorithm. The very same inputs configuration was used for the ELM whose architecture featured [15,25,45] *radial basis* neurons in the hidden layer and a linear output layer. During the initial training evaluations, the number of hidden neurons in both architectures were optimized scanning the above mentioned set of values. The resulting values were 3 and 15 respectively for SNN and ELM architecture. The first 4 weeks of recordings was extracted to provide for an initial offline calibration set, namely $T_I$, for the two ML components (SNN, ELM) to be trained. The 25% of the calibration set was used for early stopping purposes with the number of consecutive performance decreasing events empirically set at 10. After this, the two calibration functions have been adapted using different patterns of periodicity and update

(input, labels) tuples amount (see Fig. 7). Specifically, the periodicity of calibration updates τ was selected from the following set :

$$\tau \in T=[2,12,24,120,240,720,2160] \text{ hours}$$

actually ranging from two hours to three months while the amount of available updating tuple π was selected from the following set: $\pi \in P=[1,4,12,24,120,168]$ hours. Of course, only the subset of patterns in TxP featuring π < τ were plausible and hence chosen for testing purposes. Update tuple selection was made both on a regular schedule (i.e. the initial tau samples of the new T period) or uniformly sampled from the new T period tuples simulating opportunistic updates.

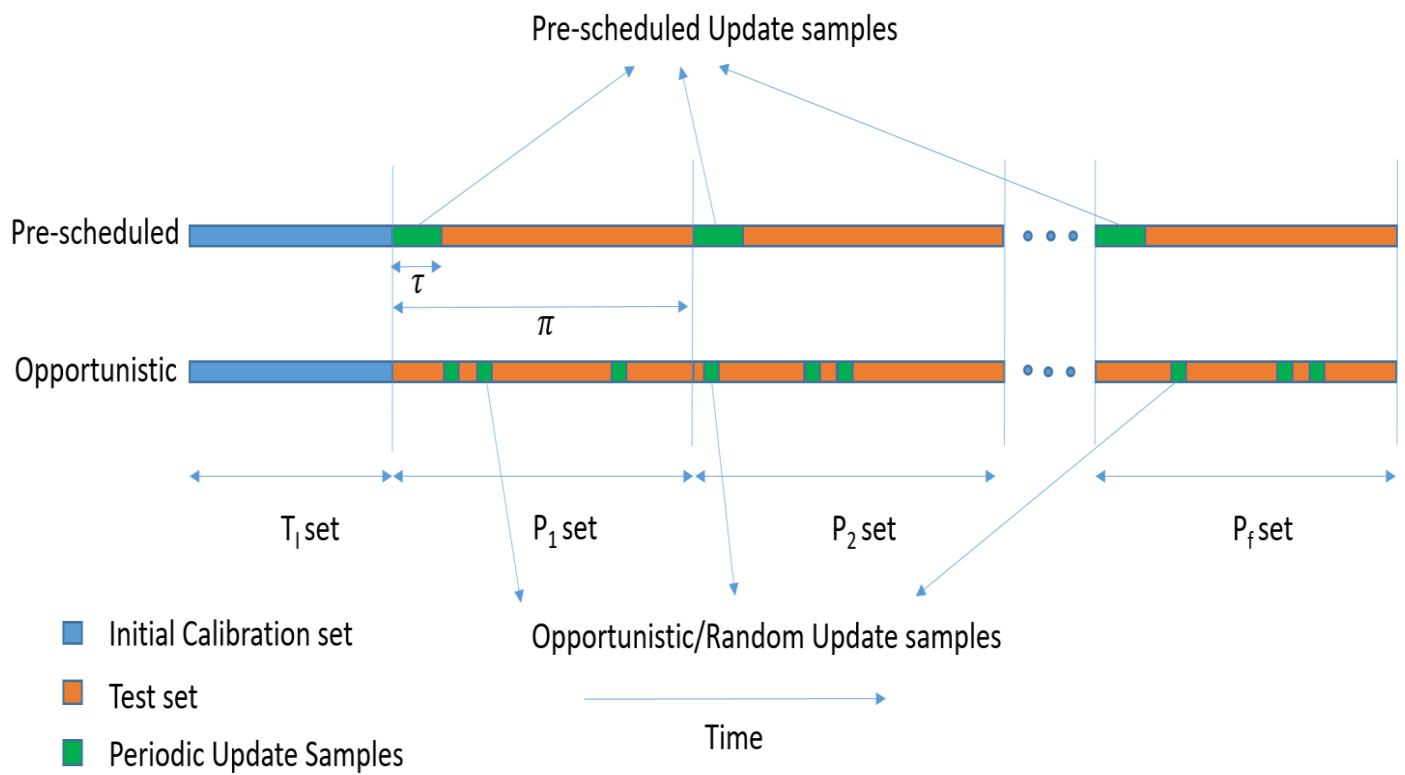

**Figure 7:** Training/Update/Test cycle procedure depicted for both regular Pre-scheduled and Opportunistic/Random update scenarios.

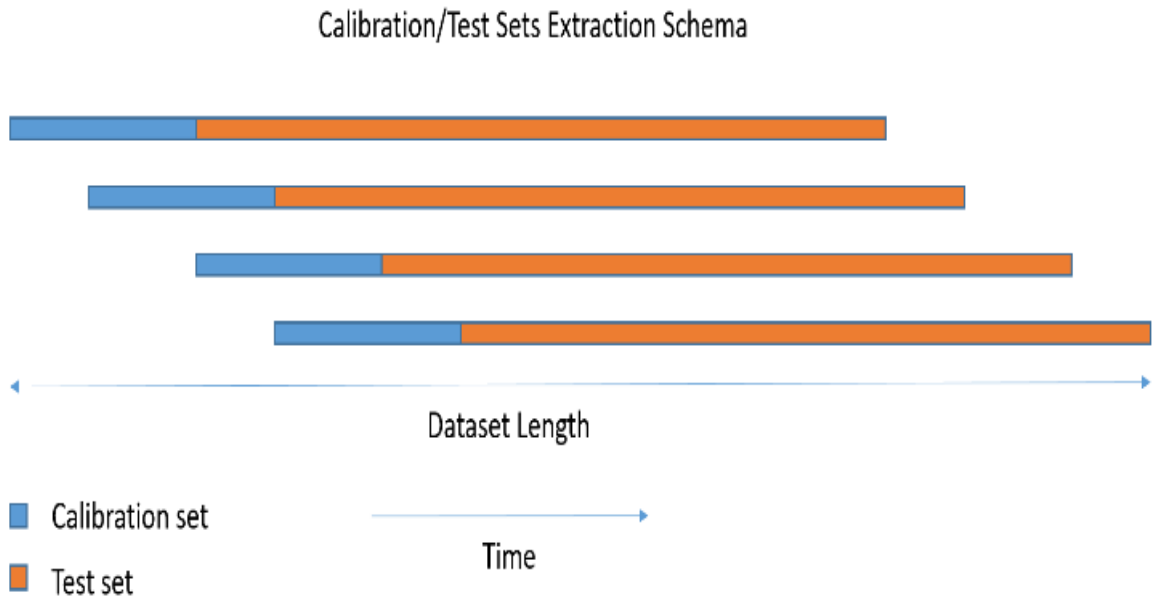

Figure 8: Validation scheme: 4 consecutive weeks have been chosen as initial training set length. Four repetitions of training/update/test procedures have been implemented, using 4 different starting time for the calibration sets. Performance indicators have been averaged among them so to reduce performance dependence from the specific sample sets choice.

Practically, the SNN was simply retrained each τ hours using the enlarging dataset (incremental approach afterwise termed iSNN) encompassing $T_I$ set plus past and newly incoming labeled tuples τ. Again, ABR was used as a training algorithm with a fixed maximum number of training epochs (500). Instead, the initially calibrated set of output weights of the ELM was *updated* using the same tuples (updating approach). Of course, ELM updating procedure provided a faster adaptive strategy with a fixed and predictable computational cost. On the other side, the SNN re-training computational cost (and hence, needed time) gradually grew with the increasing dimensions of the incremental training set. For both components, the samples of each new period where estimated using the updated calibrations outcomes and became part of the concentration estimation test data. Instead, samples up to the last incoming updating tuple concentrations were estimated by using past calibrations and integrated in the estimation test data. Estimations and true concentrations have been hence compared for performance estimation. Mean Absolute Error (MAE), Mean

Relative Error, normalized MAE (MAnE), RMSE and nRMSE have been used as performance estimators. The obtained results have been also compared with those obtained by the initial 4 weeks offline trained SNN and by a multivariate linear model trained with the same values and using ordinary least squares parameter optimization scheme. In order to reduce performance dependence from the initial 4 weeks calibration set, we repeated the experiments 4 times with different starting time after [0 week, 2 weeks, 4 weeks, 6 weeks] actually using the following 5075 samples for updates and testing purposes (see Fig. 8). 5075 is actually the largest samples size to guarantee an equally sized test set for all the calibration sets when setting it to start right after the end of the calibration sets themselves. Initial training results have been averaged 10 times for reducing dependence to the random initialization of weights in the two neural networks architectures.

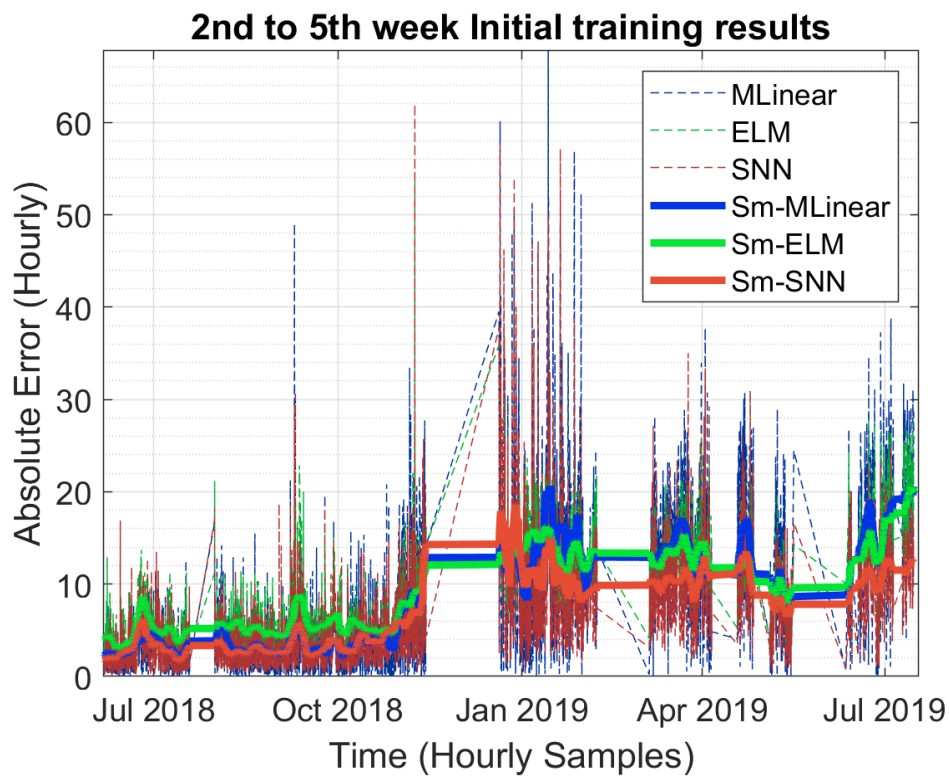

**Figure 9:** Results from one of the 4 instances of the initial training experiment for the best hyperparameter set (see supplemental material section for the remaining 3 instances). Bold lines are computed with a symmetric 96hrs long moving average smoothing filter.

## 3. Results

### 3.1. Initial Calibration Strategy

The initial calibration strategy experiment was conducted with the aim of providing insights on the basic performance level achievable with state of the art methodology.

For the best set of hyperparameters values the averaged results were reported in table 1:

**Table 1. Summary of the different performance indicator as computed after initial calibration and for the successive 5075 samples.**

|        | MAE (ppb) | MAnE  | MRE  | RMSE (ppb) | NMSE |
|--------|-----------|-------|------|------------|------|
| Linear | 7.49      | 0.075 | 0.53 | 10.44      | 0.80 |
| ELM    | 8.63      | 0.087 | 0.63 | 11.46      | 0.88 |
| SNN    | 7.43      | 0.075 | 0.53 | 10.72      | 0.82 |

Synthetic indexes indicates that performances of a multilinear and SNN models were very similar while ELM was slightly relatively underperforming. A careful analysis of the MAE performance over time clearly indicates that all the models gradually loose their validity worsening their absolute error performances with respect to the weeks immediately following the training period (see for example fig. 9 starting from October, 2018 onward).

### 3.2. Adaptive Calibration Strategies

The different TxP that have been considered are representative of very different calibration updates scenarios. In fact, (2,1) combination is representative of a deployment scenario in which the node is actually permanently co-located with a high accuracy instrument, namely a regulatory grade instrument and hence operating as a backup system for coping with the regulatory grade station faults periods. On the other extreme, the (720,168) is representative of a system that is recalibrated each month benefitting of one entire week of high accuracy calibration tuples. Intermediate scenarios like, for example,

(24,1) could be representative of model based network calibration issuing regular or, random/opportunistic (or even carefully selected based on particular updates based on particular conditions. In addition to the previous scenario, (24, 4) could be representative of a car, bus or truck mounted AQMS that is purposely benefitting of regulatory grade instrument co-location during off duty periods. Extension to on the fly node-to-node calibration scenario could and should be carefully made, due to minimum colocation period of one hour, for most of the random updates scenarios.

Table 2. Summary of the averaged MAE (ppb) results of the periodic incremental SNN (iSNN) experiments along the different TxP combinations.

|         | T=2 Hrs | T=12 Hrs | T=24 Hrs | T=240 Hrs | T=720 Hrs |
|---------|---------|----------|----------|-----------|-----------|
| P=1Hr   | 3.76    | 4.03     | 4.26     | 5.33      | 6.55      |
| P=4Hrs  | -       | 3.64     | 3.90     | 4.92      | 6.35      |
| P=12Hrs | -       | -        | 3.59     | 4.45      | 5.85      |
| P=24Hrs | -       | -        | -        | 4.27      | 5.46      |
| P=168Hrs| -       | -        | -        | 3.96      | 5.09      |
|         |         |          |          |           |           |

Tables 2 to 5 shows part of the obtained results in the different configurations in both systematic and opportunistic update scenarios, complete tables are available as supplementary materials. As a general consideration, as could be expected, the best results have been obtained in the case in which an almost continuous upgrade can be implemented. For SNN that meant a reduction of almost 50% of the MAE obtained in the initial training scenario. Similar results can be obtained starting at different T/P ratios depending on P. For example, in iSNN, a reduction of 40% can be obtained at P/T=1/24 for T=24 while a ratio of

1/5 is still not sufficient at T=720.This means that both periodicity of the retraining and the frequency of labelled sample availability are relevant for obtaining optimal performances. In general the rarer the calibration upgrade the higher the needed P/T ratio, i.e. the higher the relative number of labelled samples is needed to obtain similar performances. As the upgrades and labelled samples become more and more rare the results approach those obtainable by an initial calibration. Adaptive ELM (aELM) results are generally appreciably worse that their incremental SNN (iSNN) counterparts results but considering the similar results obtained in the initial training scenario this may well be due to a stronger representation capability expressed by the SNN architecture with respect to the ELM one.

These results are very similar but slightly better of those obtained in the opportunistic/selective update scenario when labelled sample have been made available as uniformly randomly distributed. Fig. 10 graphically shows a comparison among the different strategies results in the case of using the first 4 weeks for initial training, T=24, P=1, and systematic update. Finally, in Fig. 11, We can graphically compare the results obtained at different P/T ratio by iSNN in opportunistic and regular updates scenarios. In both cases, rare retraining (higher T) cause worsening performance even when considering same or equal incoming updates frequencies (P/T).

**Table 3. Summary of the averaged MAE (ppb) results of the periodic updated adaptive ELM (aELM) experiments along the different TxP combinations.**

|         | T=2 Hrs | T=12 Hrs | T=24 Hrs | T=240 Hrs | T=720 Hrs |
|---------|---------|----------|----------|-----------|-----------|
| P=1Hr   | 3.74    | 4.68     | 5.83     | 6.09      | 7.62      |
| P=4Hrs  | -       | 4.73     | 5.03     | 6.39      | 8.90      |
| P=12Hrs | -       | -        | 4.63     | 6.20      | 6.78      |
| P=24Hrs | -       | -        | -        | 5.71      | 6.05      |
| P=168Hrs| -       | -        | -        | 4.52      | 5.48      |

**Table 4.** Summary of the averaged MAE (ppb) results of the opportunistic incremental SNN (iSNN) experiments along the different TxP combinations.

|  | T=2 Hrs | T=12 Hrs | T=24 Hrs | T=240 Hrs | T=720 Hrs |
|---|---|---|---|---|---|
| P=1Hr | 3.74 | 4.23 | 4.72 | 6.94 | 6.56 |
| P=4Hrs | - | 3.73 | 3.97 | 5.27 | 6.31 |
| P=12Hrs | - | - | 3.65 | 4.95 | 5.73 |
| P=24Hrs | - | - | - | 4.56 | 5.40 |
| P=168Hrs | - | - | - | 3.93 | 4.99 |

**Table 5.** Summary of the averaged MAE (ppb) results of the periodic randomic update adaptive ELM (aELM) experiments along the different TxP combinations.

|  | T=2 Hrs | T=12 Hrs | T=24 Hrs | T=240 Hrs | T=720 Hrs |
|---|---|---|---|---|---|
| P=1Hr | 4.47 | 5.78 | 5.90 | 7.54 | 8.6 |
| P=4Hrs | - | 4.48 | 5.14 | 6.52 | 6.91 |
| P=12Hrs | - | - | 4.70 | 6.03 | 7.43 |
| P=24Hrs | - | - | - | 5.45 | 6.28 |
| P=168Hrs | - | - | - | 5.60 | 6.20 |

## 4. Conclusions

In this paper, we have shown the results obtainable by different calibration updates scenarios to overcome concept drift effects arising during multiseasonal deployment of low cost AQMS. With a particular focus to network calibration and using online machine learning components, we explored different combinations of updates periodicity and amount of incoming GT labelled samples. Results showed the possibility to strongly ameliorate the performance obtainable over more than one year by a field deployed AQMS receiving regular or opportunistic information from high accuracy labelled data sources.

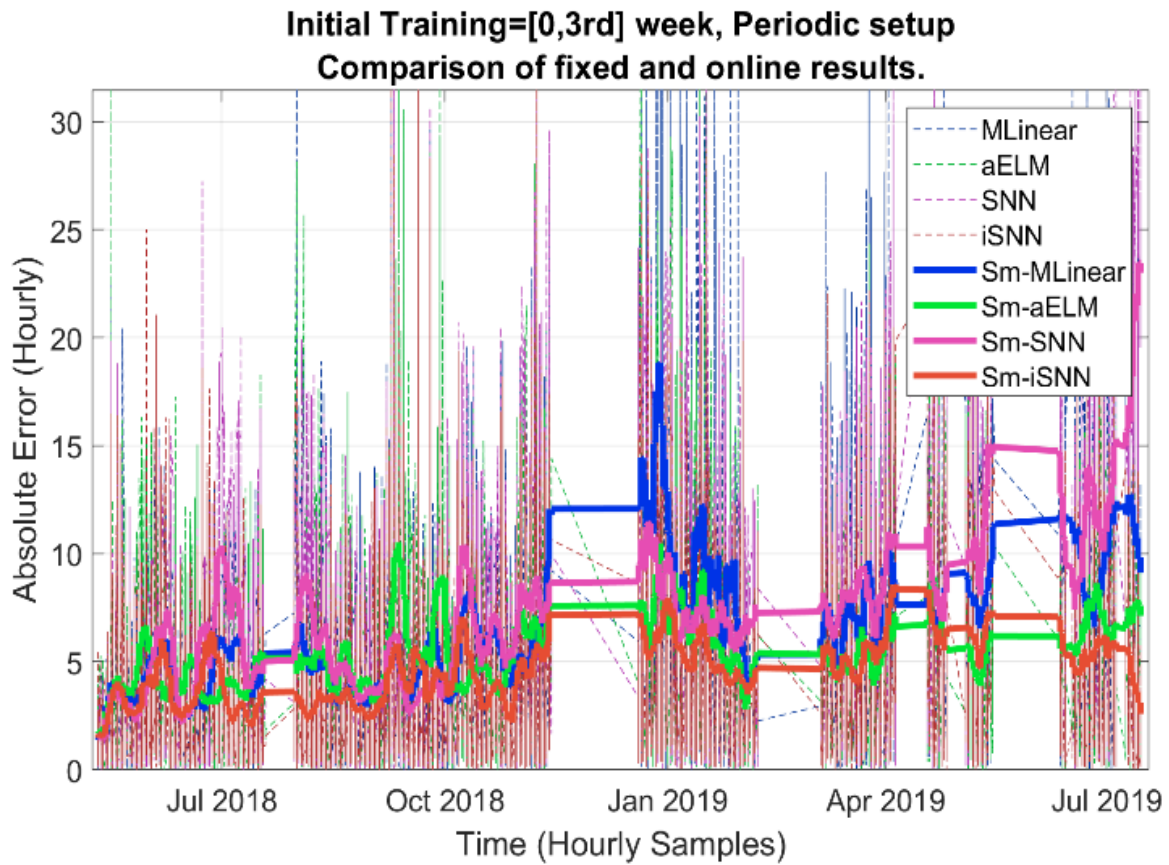

**Figure 10:** Difference among MAE performances among the different strategies as computed along the entire dataset. Note the advantages of both online learning (iSNN, aELM) vs fixed initial training approach (MLinear,SNN). Bold lines are computed with a symmetric 96hrs long moving average smoothing filter.

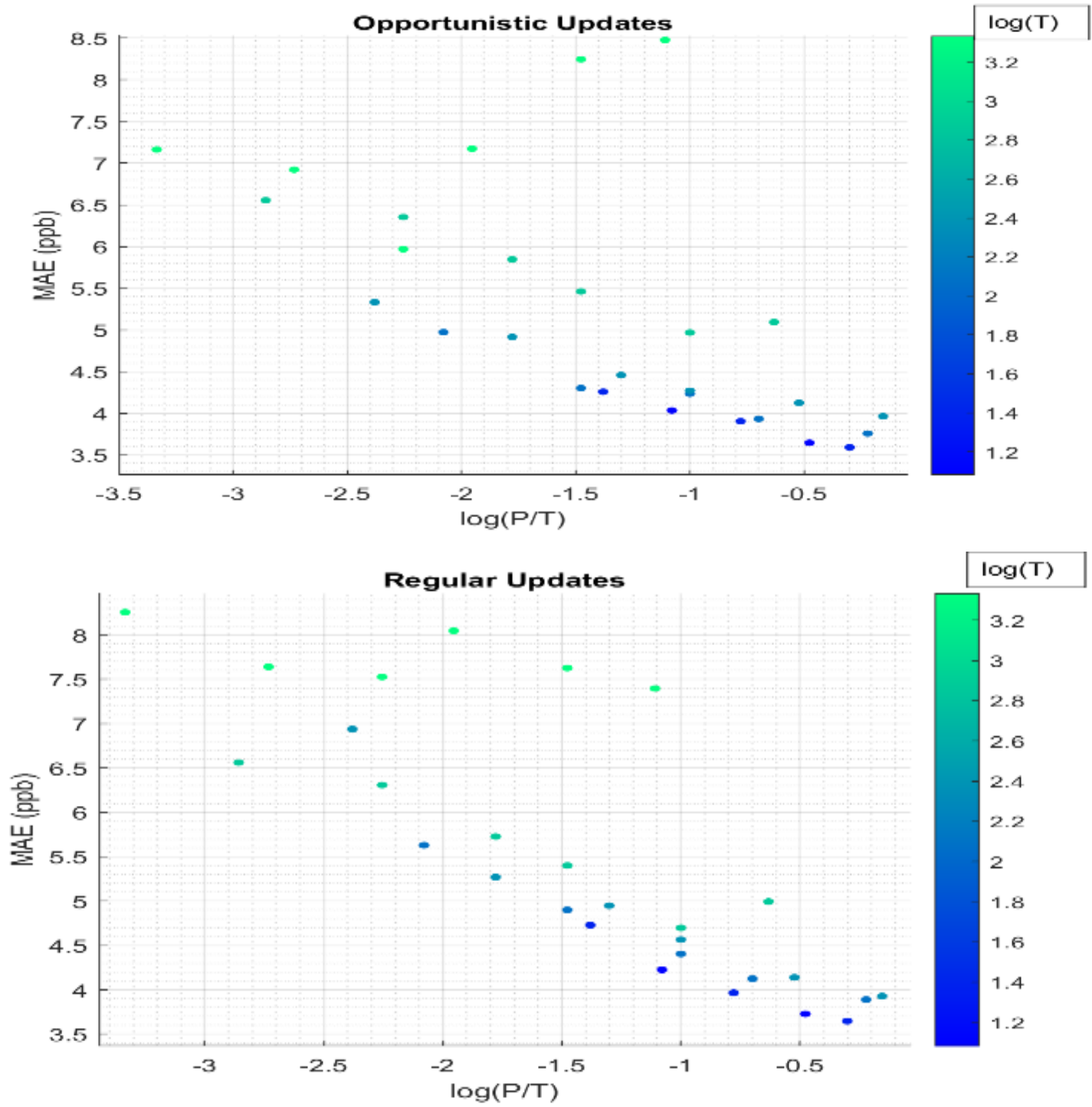

**Figure 11:** Comparison of regular and opportunistic updates strategies iSNN performances at different (T,P) combinations.

Furthermore, results indicate the relevance of both update periods and amount of labelled samples in determining the final performances. Specifically, the rarest updates requires a higher percentage amount of labelled data to reach the same preset performance goals. A probable scenario for networked calibration updates, i.e. the daily updates using a single hour of high accuracy data, can reach interesting updates closely approaching results only obtainable by continuous calibration. In order to obtain similar results a monthly recalibration may require up to one week of labelled hourly data samples. However, rarely the geostatistical models may match the accuracy of AQRMSs and further work is needed to fully understand the influence of noisy updates on the performance of periodically updated field deployed AQMS.

## Acknowledgments

This work has received partial funding by EU through UIA 3$^{rd}$ call Project AirHeritage. Authors wish to thank the Campania Regional Environmental Protection Agency (ARPAC) for continuous support in the data gathering process.

**SUPPLEMENTARY MATERIALS:**

**Table s1: Incremental SNN approach (iSNN) complete results.**

|   |     | T     |       |       |       |       |       |       |
|---|-----|-------|-------|-------|-------|-------|-------|-------|
|   |     | 2     | 12    | 24    | 120   | 240   | 720   | 2160  |
| P | 1   | 0     | 4.033 | 4.257 | 4.972 | 5.332 | 6.555 | 7.163 |
|   | 4   | 0     | 3.645 | 3.90  | 4.301 | 4.915 | 6.353 | 6.922 |
|   | 12  | 0     | 0     | 3.59  | 4.234 | 4.457 | 5.846 | 5.968 |
|   | 24  | 0     | 0     | 0     | 3.932 | 4.271 | 5.460 | 7.173 |
|   | 120 | 0     | 0     | 0     | 3.757 | 4.124 | 4.967 | 8.246 |
|   | 168 | 0     | 0     | 0     | 0     | 3.964 | 5.093 | 8.476 |

**Table s2: Adaptive ELM approach (aELM) complete results.**

|   |     | T     |       |       |       |       |       |       |
|---|-----|-------|-------|-------|-------|-------|-------|-------|
|   |     | 2     | 12    | 24    | 120   | 240   | 720   | 2160  |
| P | 1   | 0     | 4.684 | 5.829 | 5.652 | 6.094 | 7.620 | 8.069 |
|   | 4   | 0     | 4.725 | 5.037 | 5.448 | 6.391 | 8.940 | 9.089 |
|   | 12  | 0     | 0     | 4.634 | 5.661 | 6.197 | 6.782 | 9.067 |
|   | 24  | 0     | 0     | 0     | 5.311 | 5.710 | 6.056 | 6.952 |
|   | 120 | 0     | 0     | 0     | 4.734 | 5.469 | 6.008 | 6.265 |
|   | 168 | 0     | 0     | 0     | 0     | 4.525 | 5.481 | 6.652 |

**Opportunistic Update strategy complete results.**

**Table s3: Incremental SNN approach (iSNN) complete results.**

|   |     | T     |       |       |       |       |       |       |
|---|-----|-------|-------|-------|-------|-------|-------|-------|
|   |     | 2     | 12    | 24    | 120   | 240   | 720   | 2160  |
| P | 1   | 0     | 4.225 | 4.728 | 5.629 | 6.938 | 6.561 | 8.255 |
|   | 4   | 0     | 3.728 | 3.966 | 4.899 | 5.269 | 6.307 | 7.639 |
|   | 12  | 0     | 0     | 3.647 | 4.404 | 4.949 | 5.727 | 7.527 |
|   | 24  | 0     | 0     | 0     | 4.125 | 4.564 | 5.400 | 8.046 |
|   | 120 | 0     | 0     | 0     | 3.889 | 4.139 | 4.698 | 7.626 |
|   | 168 | 0     | 0     | 0     | 0     | 3.927 | 4.993 | 7.396 |

**Table s4: Adaptive ELM approach (aELM) complete results.**

|   |     | T     |       |       |       |       |       |       |
|---|-----|-------|-------|-------|-------|-------|-------|-------|
|   |     | 2     | 12    | 24    | 120   | 240   | 720   | 2160  |
| P | 1   | 0     | 5.783 | 5.904 | 7.449 | 7.536 | 8.610 | 8.832 |
|   | 4   | 0     | 4.476 | 5.147 | 6.143 | 6.523 | 6.918 | 9.727 |
|   | 12  | 0     | 0     | 4.694 | 5.265 | 6.033 | 7.438 | 9.466 |
|   | 24  | 0     | 0     | 0     | 4.966 | 5.451 | 6.286 | 8.751 |
|   | 120 | 0     | 0     | 0     | 5.381 | 4.699 | 5.996 | 8.635 |
|   | 168 | 0     | 0     | 0     | 0     | 5.634 | 6.212 | 7.385 |

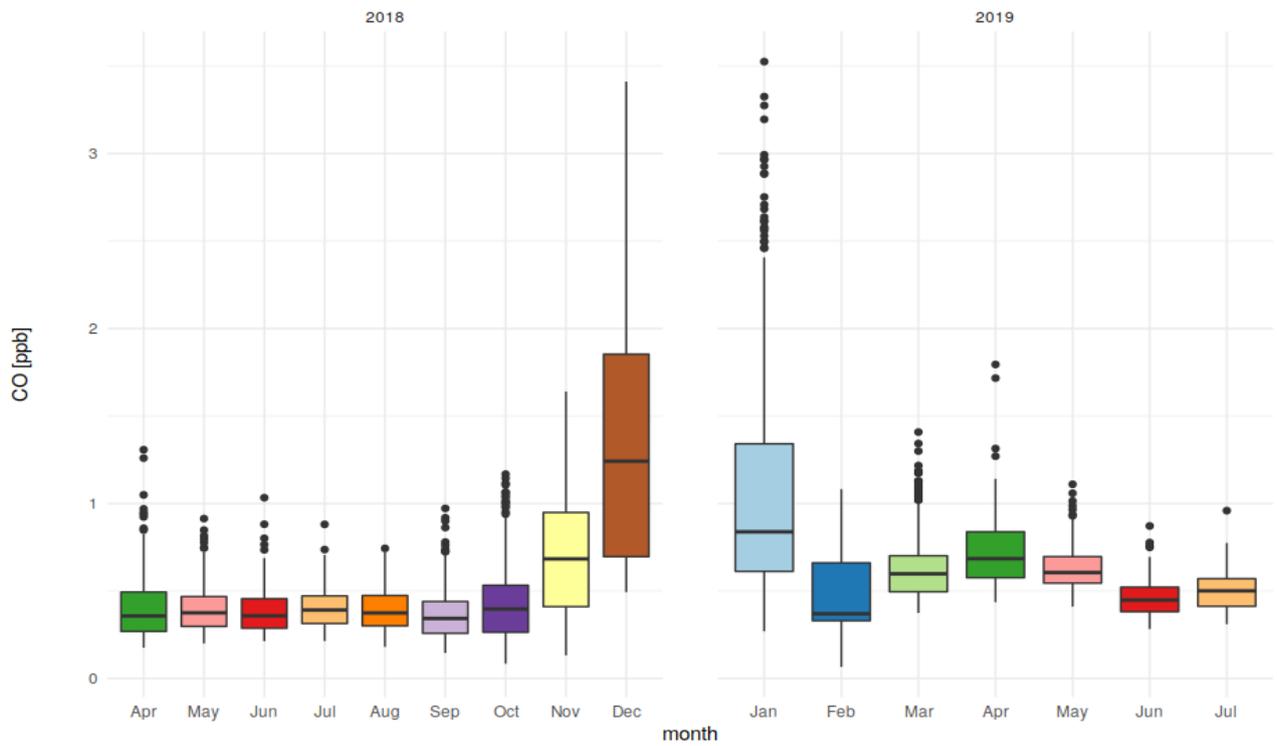

(a)

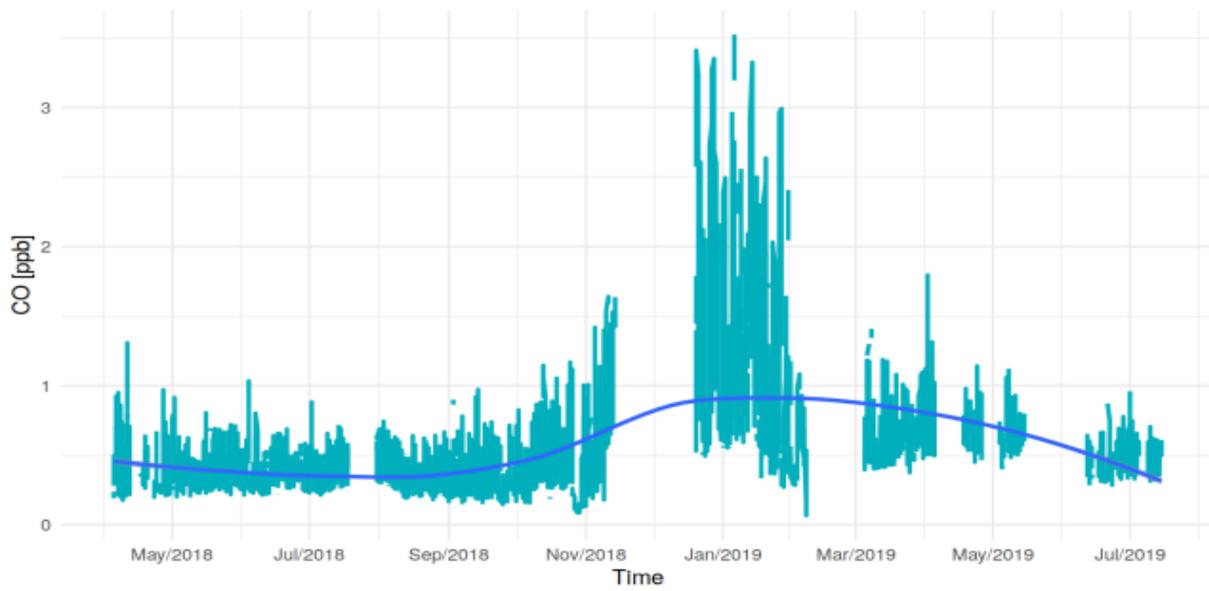

**Figure s1: Monthly box plot (a) and LOESS fitted time serie (b) of CO (ppm) reference concentration behaviour along the data set (a). Outliers are depicted as single dots.**

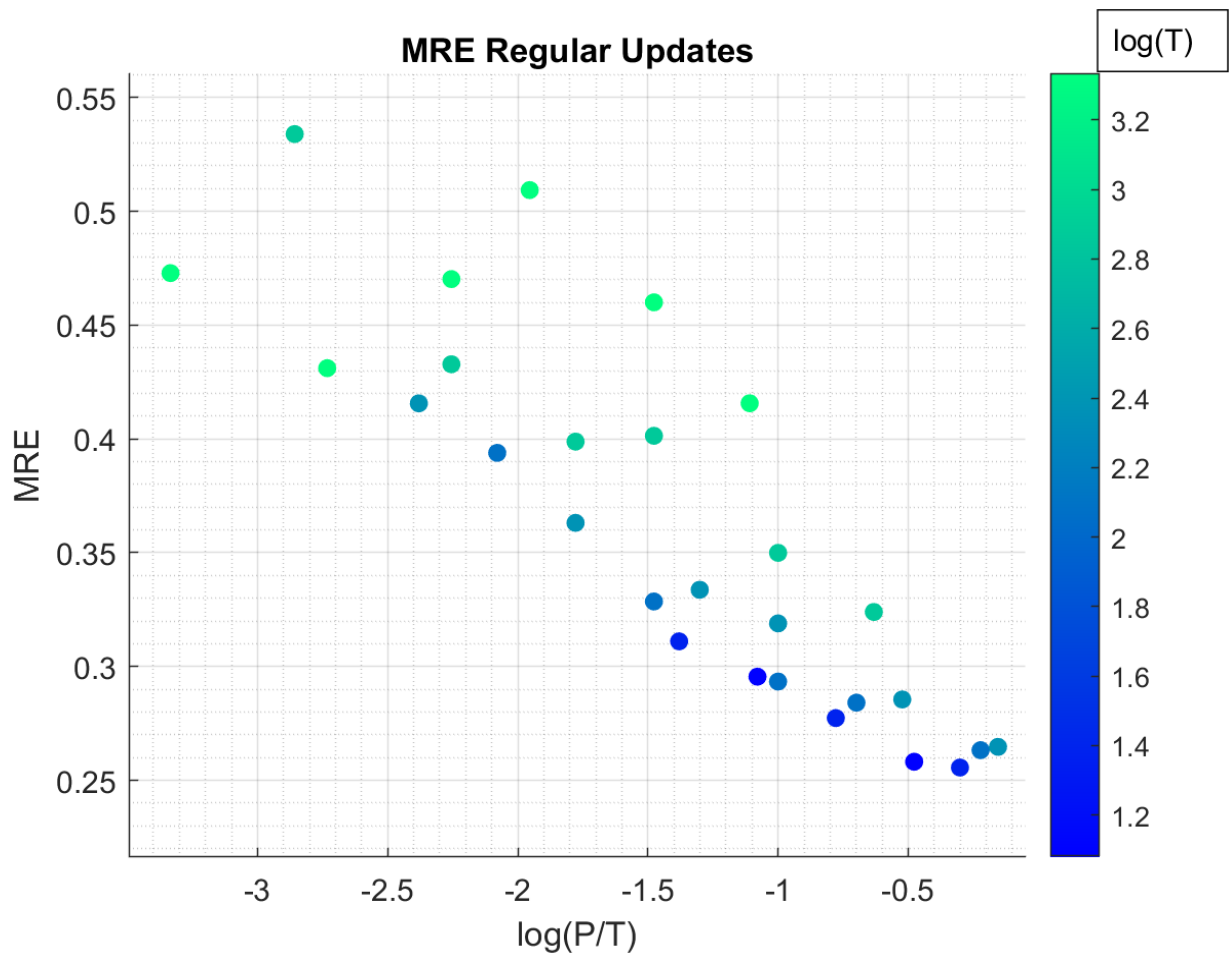

Figure s2: Semilog plot of Mean Relative Error as computed at different P/T ratios for the regular update scenario and iSNN approach. Higher ratios, when coupled with smaller T values, generate smaller MRE across the dataset duration.

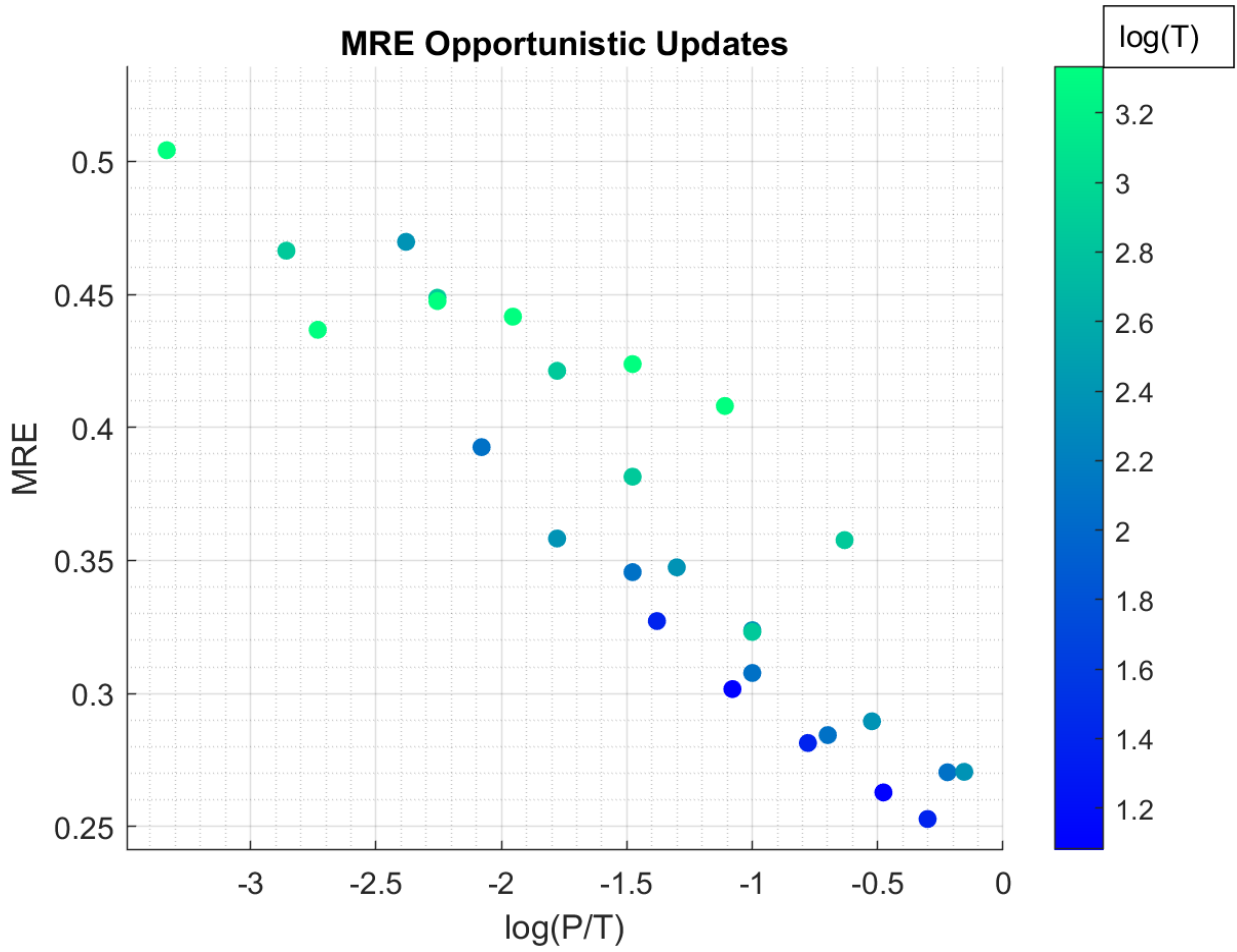

**Figure s3:** Semilog plot of Mean Relative Error as computed at different P/T ratios for the opportunistic update scenario and iSNN approach. Higher ratios, when coupled with smaller T values, generate smaller MRE across the dataset

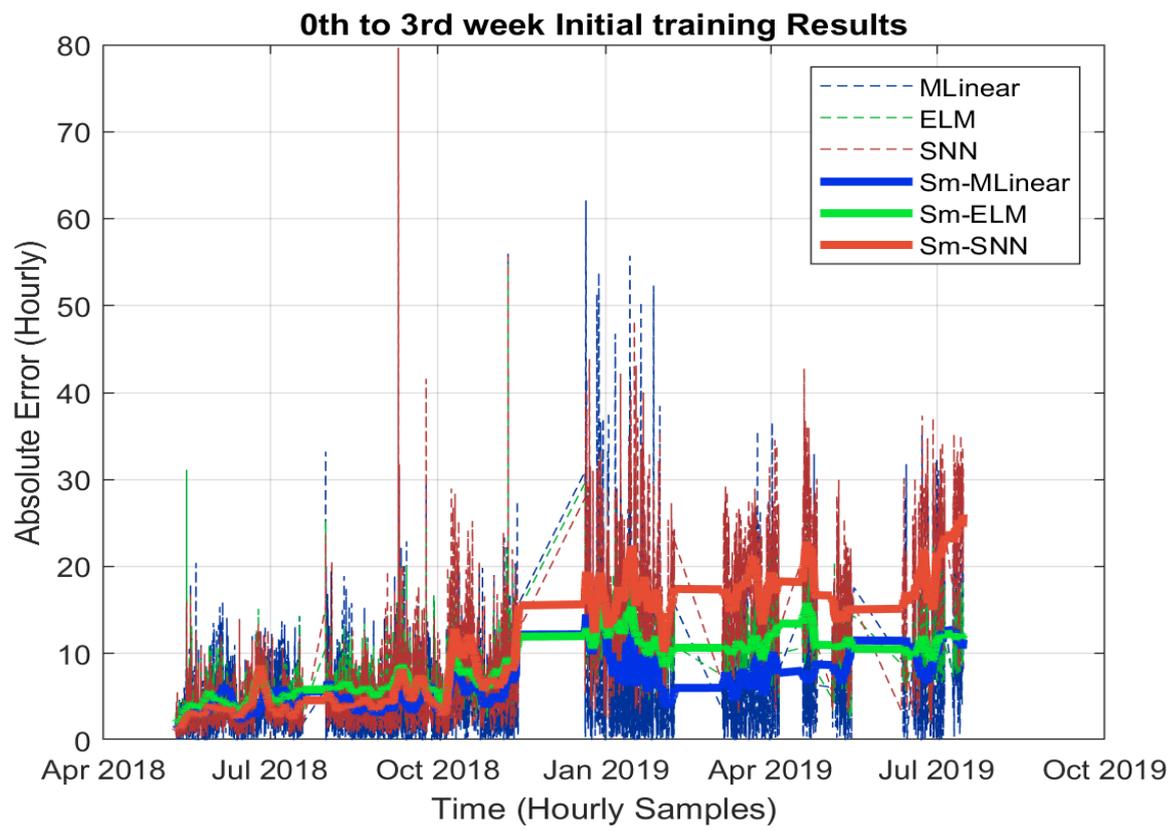

**Figure s4:** Results of the initial training experiment [0$^{th}$ to 3$^{rd}$ week) for the best hyperparameter set. Bold lines are computed with a symmetric 96hrs long moving average smoothing filter.

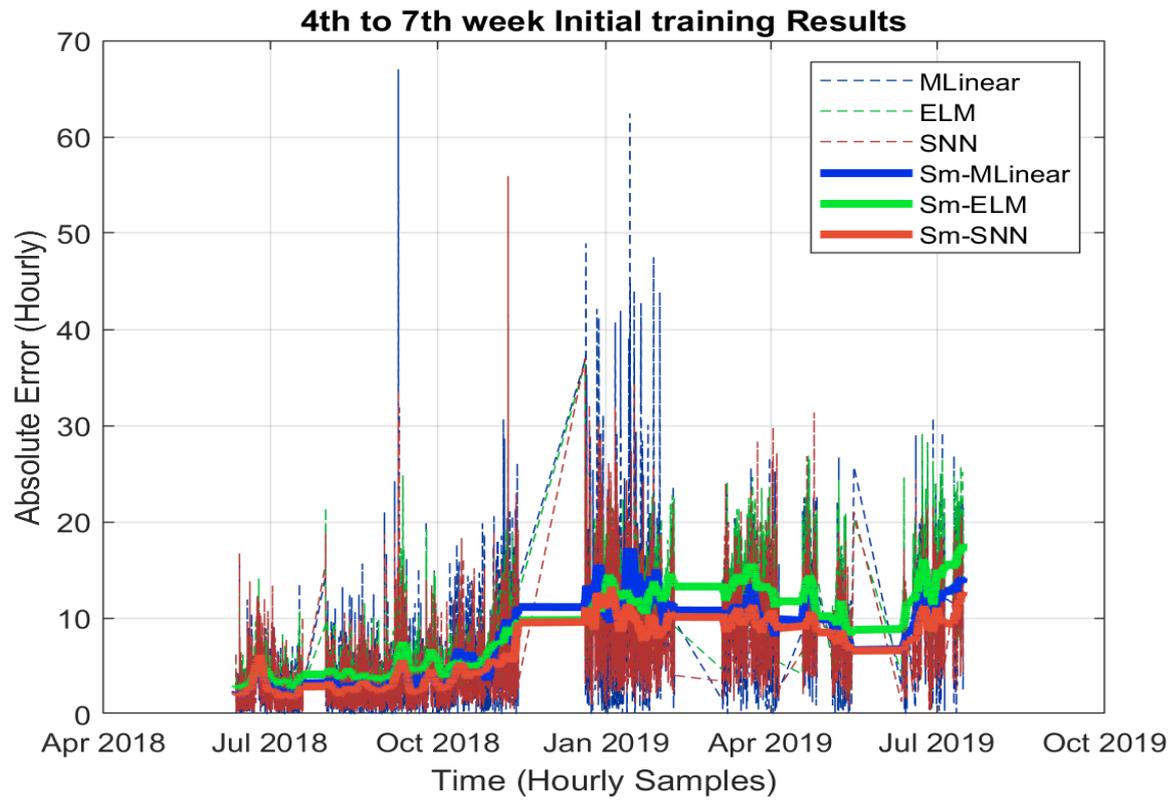

**Figure s5:** Results of the initial training experiment [4th to 7th week] for the best hyperparameter set. Bold lines are computed with a symmetric 96hrs long moving average smoothing filter.

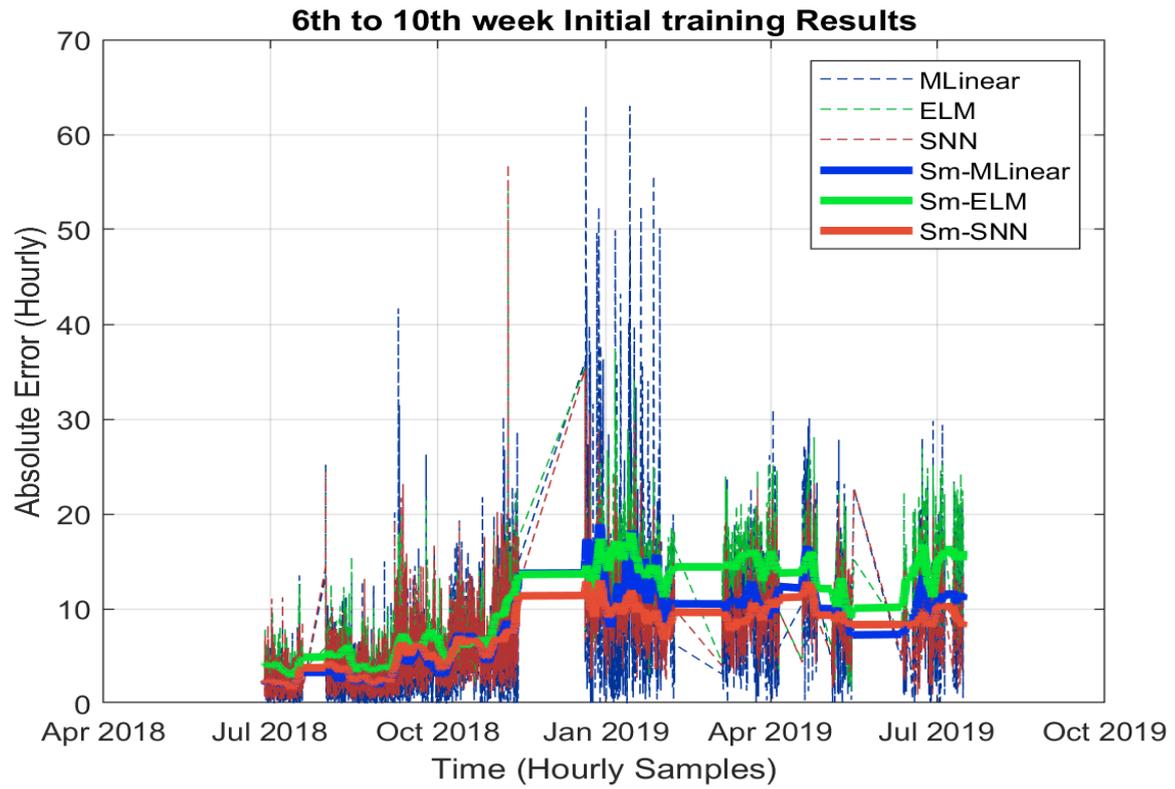

**Figure s6: Results of the initial training experiment [6th to 10th week] for the best hyperparameter set. Bold lines are computed with a symmetric 96hrs long moving average smoothing filter.**

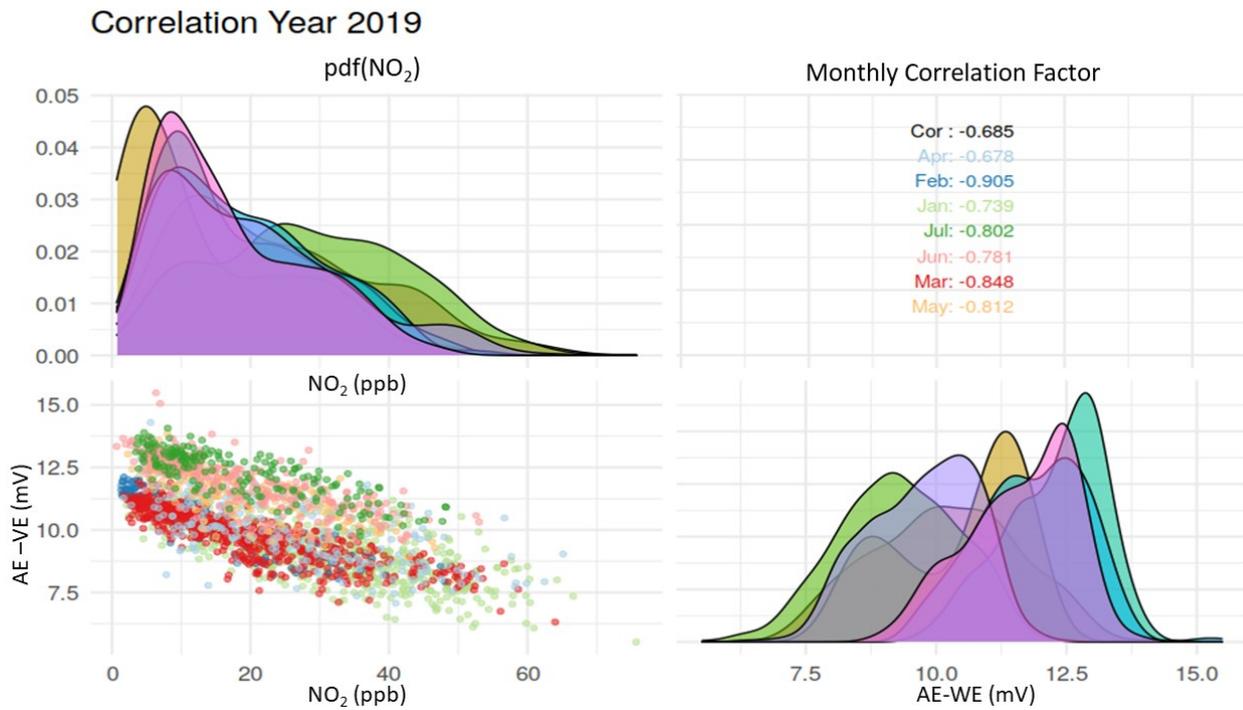

**Figure s7:** Pearson Correlation Coefficients between NO2 target gas concentration and NO2 raw sensor readings (AE-WE difference) as computed for each month in 2019 (upper right corner), monthly empirical pdf (upper left corner for NO2 and lower right for AE-WE) and scatter plot among the above variables (lower left corner). Different colors are associated to different monhts. Significant, seasonal bias and sensitivity changes are evident.